\documentclass[prb,preprint,draft,amsmath,showpacs]{revtex4}

\usepackage{graphicx}
\begin{document}

\DeclareGraphicsExtensions{.eps, .jpg}

%%%%%%%%%%%%%%%%%%%%%%%%%%%%%%%%%%%%%%%%%%

\bibliographystyle{prsty}
\input epsf

\title {Optical investigation of the metal-insulator transition in $FeSb_2$}

\author {A. Perucchi$^{1}$, L. Degiorgi$^{1}$, R. Hu$^{2,3}$, C. Petrovic$^{2}$ and V. Mitrovi{\'c}$^{3}$}
\affiliation{$^{1}$Laboratorium f\"ur Festk\"orperphysik, ETH Z\"urich,
CH-8093 Z\"urich, Switzerland}\
\affiliation{$^{2}$Physics Department, Brookhaven National Laboratory, Upton NY 11973, U.S.A.}\
\affiliation{$^{3}$Physics Department, Brown University, Providence RI 02912, U.S.A.}\
\date{\today}

\begin{abstract}
We present a comprehensive optical study of the narrow gap $FeSb_2$ semiconductor. From the optical reflectivity, measured from the far infrared up to the ultraviolet spectral range, we extract the complete absorption spectrum, represented by the real part $\sigma_1(\omega)$ of the complex optical conductivity. With decreasing temperature below 80 K, we find a progressive depletion of $\sigma_1(\omega)$ below $E_g\sim 280$ cm$^{-1}$, the semiconducting optical gap. The suppressed (Drude) spectral weight within the gap is transferred at energies $\omega>E_g$ and also partially piles up over a continuum of excitations extending in the spectral range between zero and $E_g$. Moreover, the interaction of one phonon mode with this continuum leads to an asymmetric phonon shape. Even though several analogies between $FeSb_2$ and $FeSi$ were claimed and a Kondo-insulator scenario was also invoked for both systems, our data on $FeSb_2$ differ in several aspects from those of $FeSi$. The relevance of our findings with respect to the Kondo insulator description will be addressed.
\end{abstract}
\pacs{78.20.-e}
\maketitle

\section{Introduction}
$FeSb_2$ represents an interesting case of narrow gap semiconductor, where a band of itinerant electron states originates in the $d_{xy}$ orbitals of the $t_{2g}$ multiplet. The magnetic properties \cite{petrovic03,petrovic05} of $FeSb_2$ strongly resemble those seen in $FeSi$ (Refs. \onlinecite{jaccarino67} and \onlinecite{mandrus95}). The magnetic susceptibility of $FeSb_2$ shows a diamagnetic to paramagnetic crossover around 100 K (Ref. \onlinecite{petrovic03}); however in contrast to $FeSi$, with a very small low temperature ($T$) impurity tail in the diamagnetic region. The electrical transport in $FeSb_2$ along two axes is semiconducting, whereas the third axis exhibits a metal-semiconductor crossover at temperature $T_{cr}\sim 40-80$ K, depending from the current alignment \cite{petrovic03}. Furthermore, the thermal expansion and heat capacity of $FeSb_2$ at ambient pressure agrees with a picture of a temperature induced spin state transition within the $Fe$ $t_{2g}$ multiplet \cite{petrovic05}.

Aeppli and Fisk \cite{aeppli92} proposed that the underlying physics of $FeSi$ share common features with a class of rare earth compounds, known as hybridization gap semiconductors or Kondo insulators. The analogies between $FeSi$ and $FeSb_2$, pointed out above, would suggest that also $FeSb_2$ might belong to the class of $d$-electron based Kondo insulators. Within the Kondo insulator's scenario, one considers two narrow hybridized bands of width $W$ in the density of states, separated by the gap $E_g$. At $T=0$ the electrons populate a lower hybridized band, and with the increase of $T$ the electrons start populating the higher band, resulting in a thermally activated Pauli susceptibility \cite{aeppli92}. This approach, successfully applied to $FeSi$, seems to work out for $FeSb_2$, as well \cite{petrovic03,petrovic05}. This is of particular interest because of possible relations between theoretical descriptions of $d$- and $f$-electron systems.

$FeSi$ was also intensively investigated from the point of view of its absorption spectrum \cite{degiorgiepl,chernikov97,paschen97,schlesinger93}. The charge excitation spectrum of $FeSi$ is characterized at low temperatures by a direct gap of about 95 meV. The spectral weight, suppressed by the gap opening, is transferred to high energies. It has been long debated up to which energy the total spectral weight in $FeSi$ is effectively fully recovered \cite{degiorgiepl,schlesinger93}. In the data of Degiorgi {\it et al.} (Ref. \onlinecite{degiorgiepl}), the spectral weight is essentially recovered at a frequency $\omega_c\sim 4E_g$, therefore without any need to invoke an integration of $\sigma_1(\omega)$ to very high frequencies. This conclusion is at variance with claims \cite{schlesinger93}, suggesting a redistribution of spectral weight extending up to very high energies, even beyond the highest energy limit of that experiment. This feature was also found to be indicative for a Kondo insulator description of $FeSi$ (Ref. \onlinecite{schlesinger93}). As far as $FeSb_2$ is concerned, a comprehensive optical study is still missing.

In addition to $FeSi$, $FeSb_2$ seems to be a promising model system in order to investigate the applicability of the Kondo insulator concept with respect to the nearly itinerant magnetic semiconductor picture for $3d$ intermetallic compounds. The absorption spectrum of $FeSb_2$ should reveal important information about its intrinsic physical properties and should allow to extract the phonon spectrum as well as the relevant energy scales, like the hybridization gap. To test the applicability of the Kondo insulating concept, it is of interest to investigate the redistribution of spectral weight above and below $T_{cr}$ in $FeSb_2$, addressing the issue of its conservation.

We report on our optical investigation on $FeSb_2$. The paper is organized as follows: we will first characterize the specimen and describe the experiment, followed by the presentation of the optical data. The discussion will first tackle the analysis of the phonon spectrum. Particular attention will be then devoted to the issue of the spectral weight distribution. The relevance of the Kondo insulator scenario for the description of the metal-insulator transition at $T_{cr}$ in $FeSb_2$ will be emphasized.

\section{Experiment and results}
%<<<<<<<<<<<<<<<<<<<<<<<< FIGURE>>>>>>>>>>>>>>>>>>>>>>>>>
\begin{figure}[tbp]
   \begin{center}
   %\resizebox{7.0 cm}{!}{\includegraphics{Fig0.pdf}}
    \leavevmode
    \epsfxsize=10cm\epsfbox {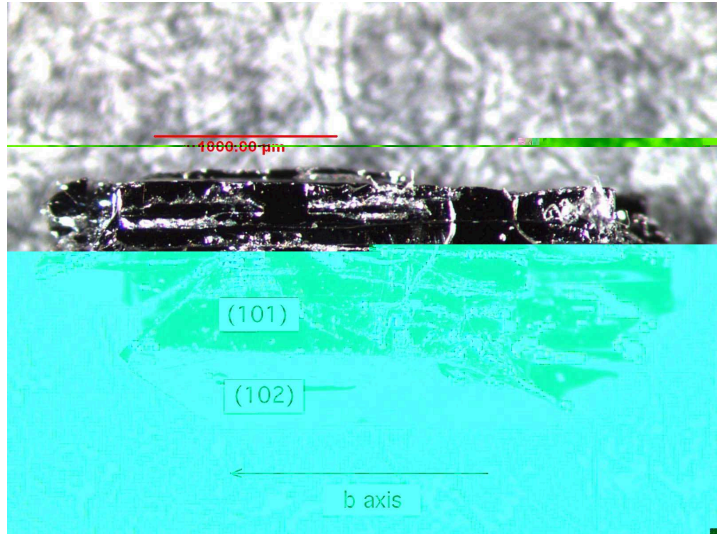}
     \caption{$FeSb_2$ sample used for our optical investigations. The crystal axes, determined through the Laue spectroscopic experiment, are also shown.}
\label{fig0}
\end{center}
\end{figure}
%<<<<<<<<<<<<<<<<<<<<<<<< figure >>>>>>>>>>>>>>>>>>>>>>>>>
Single crystals of $FeSb_2$ are grown from excess $Sb$ flux, as described in Ref. \onlinecite{petrovic03}. 
The crystals are cut in a rectangular shape with the long direction corresponding to the $b$ crystalline axis. Our sample is shown in Fig. 1, which also displays the crystal orientations. We measured the optical reflectivity $R(\omega)$ of our $FeSb_2$ specimen, over a broad spectral range (30 to 10$^5$ cm$^{-1}$) and as a function of temperature. Light was polarized along the $b$ axis ($E\parallel b$) and along the direction perpendicular ($E\perp b$) to the $b$ axis within the (102) surface (Fig. 1). The real part $\sigma_1(\omega)$ of the optical conductivity was obtained through Kramers-Kronig (KK) transformation \cite{wooten,dressel} of $R(\omega)$. To this end, we applied standard high-frequency extrapolations $R(\omega)\sim\omega^{-s}$ (with $2< s<4$) in order to extend the data set above 10$^5$ cm$^{-1}$ into the electronic continuum. Below 30 cm$^{-1}$, $R(\omega$) was extrapolated towards zero frequency either by using the Hagen-Rubens (HR) law $R(\omega$)=$1-2\sqrt{(\omega/\sigma_{dc}})$ for reflectivity data at $T>T_{cr}$, displaying a metallic behavior, or by imposing $R(\omega<\omega_{min}$)=$R(\omega_{min}=30$ cm$^{-1}$) for data in the insulating state at $T<T_{cr}$. Nevertheless, the overall behaviour of $\sigma_1(\omega)$ is not affected by the details of these extrapolations. Details pertaining to the experiment can be found in Refs. \onlinecite{wooten} and \onlinecite{dressel}.

%<<<<<<<<<<<<<<<<<<<<<<<< FIGURE>>>>>>>>>>>>>>>>>>>>>>>>>
\begin{figure}[tbp]
   \begin{center}
   %\resizebox{9.0 cm}{!}{\includegraphics{Figura1.epsf}}
    \leavevmode
    \epsfxsize=10cm \epsfbox {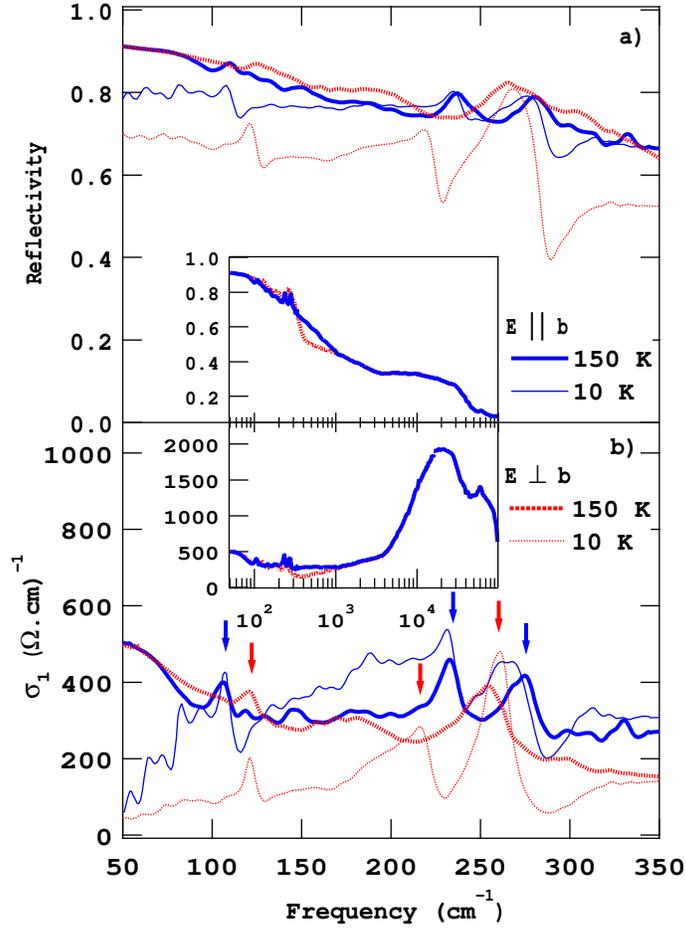}
     \caption{ (color online) a) Optical reflectivity $R(\omega)$ and b) real part $\sigma_1(\omega)$ of the optical conductivity of $FeSb_2$ in the far infrared range, with light polarized along the $b$ crystallographic axis ($E\parallel b$) and perpendicular to the $b$ axis ($E\perp b$) at 150 and 10 K. Insets: $R(\omega$) and $\sigma_1(\omega)$ for frequencies up to $10^5$ cm$^{-1}$ (the reader should note the use of the logarithmic energy scale). Blue and red arrows indicate the low temperature phonon modes for $E \parallel b$ and $E \perp b$, respectively.}
\label{fig1}
\end{center}
\end{figure}
%<<<<<<<<<<<<<<<<<<<<<<<< figure >>>>>>>>>>>>>>>>>>>>>>>>>

Figure 2 presents $R(\omega)$ and $\sigma_1(\omega)$ in the far-infrared (FIR) spectral range at 10 and 150 K for both polarizations of light, while the insets display the spectra at 150 K for both quantities over the whole measured spectral range (the reader should pay attention to the use of the logarithmic energy scale in the insets). $R(\omega$) at 150 K displays a metallic behavior for both polarization directions, with a plasma edge onset at $\sim 3000$ cm$^{-1}$ (inset of Fig. 2a). The anisotropy in $R(\omega$) is not very pronounced at 150 K, even though the $E \perp b$ reflectivity is depleted with respect to $R(\omega)$ along the $b$ axis in the spectral range between 300 and 1000 cm$^{-1}$. With decreasing $T$ and for both polarizations, $ R(\omega$) flattens in the far infrared range (Fig. 2a), an indication for a metal to insulator (MI) transition. The enhancement of phonon-like structures, especially for $E \perp b$, is also observed at 10 K (i.e., in the insulating state). 
$R(\omega$) at 150 and 10 K merge together at $\sim 350$ and 450 cm$^{-1}$, for light polarization $E \parallel b$ and $E \perp b$ , respectively. The anisotropic behavior in the optical response between the two polarization directions becomes more evident at low temperatures. In the far infrared range, $R(\omega$) at 10 K along the $b$ axis is approximately 10\% higher with respect to the other crystallographic axis. Furthermore, the anisotropic electrodynamic response clearly shows the presence of different phonon features along the two crystallographic directions.

The real part $\sigma_1(\omega)$ of the optical conductivity, as extracted from the KK transformations, is shown in Fig. 2b.  At 150 K (inset of Fig. 2b), one can easily recognize the presence of a Drude metallic term at low frequencies, for both polarization directions, as well as the strong absorption at $\sim 20000$ cm$^{-1}$ due to an electronic interband transition. With decreasing temperature, the Drude term vanishes along both directions (main panel of Fig. 2b).  Furthermore, for $E\parallel b$ we can clearly distinguish phonon modes at $\sim 107$, 231, 270 cm$^{-1}$ (blue arrows in Fig. 2b); the latter one seems to be made up by two distinct modes, yet not completely resolved. On the other hand, $\sigma_1(\omega)$ for $E \perp b$ displays phonon features at $\sim 122$, 217, 260 cm$^{-1}$ (red arrows in Fig. 2b). The mode at 217 cm$^{-1}$ is clearly present at low temperatures only.

\section{Discussion}
\subsection{The fitting procedure}

%<<<<<<<<<<<<<<<<<<<<<<<< TABLE >>>>>>>>>>>>>>>>>>>>>>>>>
\begin{table}[!ht]
\begin{centering}
  \caption{\label{P90} $FeSb_2$ - $E\parallel b$.
Lorentz-Drude fit parameters. All parameters are given in units of [cm$^{-1}$], with the exception of 
$\epsilon_{\infty}$ which is unit-less.}
\begin{ruledtabular}
   \begin{tabular}{cllllllllllll}
    \bf Parameters&\bf 10 K&\bf 30 K&\bf 50 K&\bf 70 K&\bf 100 K&\bf 150 K\\
    \hline
    \hline
    $\boldsymbol 
\epsilon_{\infty}$&1&1&1&1&1&1\\
        \hline
        \hline
$\boldsymbol \omega_{pD}$ & - & - & - & - & 1400 & 1900\\
$\boldsymbol \gamma_{D}$ & - & - & - & - & 75 & 85\\
       \hline
       \hline
$\boldsymbol \omega_{p1}$ & 580 & 510 & 480 & 480 & 300 & 300\\
$\boldsymbol \omega_{01}$ & 106.4 & 107.5 & 107.5 & 107.5 & 107.5 & 106.3\\
$\boldsymbol \gamma_1$ & 13 & 12 & 10.5 & 10.5 & 10.5 & 12\\
       \hline
$\boldsymbol \omega_{p2}$ & 650 & 650 & 650 & 680 & 590 & 560\\
$\boldsymbol \omega_{02}$ & 231 & 231 & 231 & 231 & 232 & 232.9\\
$\boldsymbol \gamma_2$  & 15 & 15.5 & 15.5 & 15.5 & 15.5 & 15\\
       \hline
$\boldsymbol \omega_{p3}$ & 635 & 600 & 615 & 520 & 520 & 455\\
$\boldsymbol \omega_{03}$ & 257 & 257 & 256.5 & 258.3 & 258.3 & 264\\
$\boldsymbol \gamma_3$  & 25 & 25 & 25 & 25 & 25 & 25\\  
       \hline
$\boldsymbol \omega_{p4}$ & 655 & 655 & 650 & 715 & 630 & 625\\
$\boldsymbol \omega_{04}$ & 271 & 272.5 & 273 & 272 & 274 & 276\\
$\boldsymbol \gamma_4$  & 25 & 23 & 20 & 22 & 22 & 25\\       \end{tabular}
\end{ruledtabular}
\end{centering}
\end{table}
%<<<<<<<<<<<<<<<<<<<<<<<< TABLE >>>>>>>>>>>>>>>>>>>>>>>>>

%<<<<<<<<<<<<<<<<<<<<<<<< TABLE >>>>>>>>>>>>>>>>>>>>>>>>>
\begin{table}[!ht]
\begin{center}
  \caption{\label{P0} $FeSb_2$ - $E\perp b$.
Lorentz-Drude and Fano fit parameters. All parameters are given  in units of [cm$^{-1}$], with the exception of
$\epsilon_{\infty}$ and $q$ which is unit-less.}
\begin{ruledtabular}
   \begin{tabular}{cllllllllllll}
    \bf Parameters&\bf 10 K&\bf 30 K&\bf 50 K&\bf 70 K&\bf 100 K&\bf 150 K\\
    \hline
    \hline
    $\boldsymbol 
\epsilon_{\infty}$&1&1&1&1&1&1\\
        \hline
        \hline
$\boldsymbol \omega_{pD}$ & - & - & - & 900 & 1484.5 & 2100\\
$\boldsymbol \gamma_{D}$ & - & - & - & 80 & 112.8 & 130\\
       \hline
       \hline
$\boldsymbol \omega_{p1}$ & 365 & 350 & 335 & 275 & 225 & 230\\
$\boldsymbol \omega_{01}$ & 121 & 121 & 122.5 & 122 & 123 & 121\\
$\boldsymbol \gamma_1$ & 11.5 & 12 & 12 & 9 & 10 & 12\\
       \hline
$\boldsymbol \omega_{p2}$ & 420 & 410 & 400 & 420 & - & -\\
$\boldsymbol \omega_{02}$ & 216 & 217 & 219 & 219.5 & - & -\\
$\boldsymbol \gamma_2$  & 13 & 13 & 13 & 18 & - & -\\  
      %\hline
      \hline
 %\bf Fano &\bf 10 K&\bf 30 K&\bf 50 K&\bf 70 K&\bf 100 K&\bf 150 K\\
% \hline
      \hline
$\boldsymbol \omega_{p3}$ & 776.1 & 746.3 & 724.4 & 722.6 & 777 & 800\\
$\boldsymbol \omega_{03}$ & 261.4 & 262.5 & 263.8 & 263.4 & 261.4 & 254.0\\
$\boldsymbol \gamma_3$  & 21.1 & 19.5 & 19.6 & 20.3 & 33.0 & 40.0\\
$\boldsymbol q_3$  & -6.41 & -6.87 & -7.67 & -7.91 & -16 & -16\\       
\end{tabular}
\end{ruledtabular}
\end{center}
\end{table}
%<<<<<<<<<<<<<<<<<<<<<<<< TABLE >>>>>>>>>>>>>>>>>>>>>>>>>

The optical conductivity is analyzed within the framework of the phenomenological Lorentz-Drude \cite{dressel,wooten} model extended to the Fano approach \cite{calvani01,fano,damascelli} in order to account for the asymmetric shape in the phonon modes. The Drude metallic behavior is defined by the formula:
\begin{equation}
\sigma_D(\omega)=\frac{\omega_{pD}^2}{4\pi\gamma_D}\frac{1}{1-i\omega/\gamma_D},
\end{equation}
where the plasma frequency is $\omega_{pD}=\sqrt{4\pi Ne^2/m}$ and $\gamma_D$ is the scattering rate. Besides the Drude term, we add four and two Lorentz harmonic oscillators for $E\parallel b$ and $E\perp b$, respectively. The contribution due to a Lorentz harmonic oscillator is:
\begin{equation}
\sigma_{Lorentz}^j(\omega)=\frac{\omega_{pj}^2}{4\pi}\frac{\omega}{i(\omega_{0j}^2-\omega^2)+\omega\gamma_j},
\end{equation}
where $\omega_{0j}$ is the resonance frequency, $\gamma_j$ the damping and $\omega_{pj}$ the strength of the $j$-mode. For $E\perp b$, there is additionally a third contribution, described by a Fano lineshape for the asymmetric mode at 260 cm$^{-1}$. The contribution due to the Fano mode is given by the equation \cite{fano,damascelli}:

\begin{equation}
\sigma_{Fano}^j(\omega)=i\sigma_{0j}\frac{(q_j+i)^2}{x_j(\omega)+i},
\end{equation}
with
\begin{equation}
x_j(\omega)=\frac{\omega_{0j}^2-\omega^2}{\gamma_j\omega}, \textrm{ and } \sigma_{0j}=\frac{\omega_{pj}^2}{\gamma_j q_j^2},
\end{equation}
where the same notation as in eq. (2) has been used. The dimensionless $q_j$ Fano parameter is a measure of the degree of asymmetry of the peak at $\omega_{0j}$  (for $|q| \rightarrow \infty$ a Lorentzian lineshape is recovered). The Lorentz and Fano contributions describe the modes, marked by arrows in Fig. 2b, which clearly develop at low temperatures.

Table I and II summarize the temperature dependence of the parameters employed in our fits. We immediately observe that with decreasing temperature, the plasma frequency decreases for both polarizations. From our fits we can define two intervals of temperatures for the MI transition, T$_{ MI}^{\parallel b}=$ 70-100 K and T$_{ MI}^{\perp b}=$ 50-70 K, as the range of temperatures at which the Drude term vanishes. Our $T_{MI}$ values are in good agreement with $T_{cr}$, which defines the metal-semiconducting crossover in the $\rho(T)$ transport data \cite{petrovic03}. Another interesting aspect in the infrared spectra of $FeSb_2$ is the quite large mode strength of the infrared active phonons. Particularly, the phonons at $\omega_{02}$, $\omega_{03}$ and $\omega_{04}$ for $E\parallel b$ and the Fano mode at $\omega_{03}$ for $E\perp b$ exhibit at low temperatures mode strengths between 600 and 800 cm$^{-1}$. This could suggest highly ionic character for $FeSb_2$. In this respect, $FeSb_2$ shares a common property with $FeSi$ (Ref. \onlinecite{schlesinger93}). Schlesinger {\it et al.} also pointed out that strong phonons have been observed in prototype heavy fermions systems and cuprates, as well \cite{schlesinger93}.

While the temperature dependence of the phonon modes and in general of $\sigma_1(\omega)$ in the far infrared (FIR) spectral range will be extensively discussed below, we anticipate here that the proposed fit only partially reproduces the measured spectra. Indeed, there is a broad continuum of low frequency excitations in the FIR spectral range and at low temperatures for both polarizations, which is not encountered by this fit procedure.

\subsection{Lattice Dynamics}
\subsubsection{Factor group analysis}
$FeSb_2$ crystallizes in the marcasyte-type structure with two formula units per unit cell. $FeSb_2$ is made up of $FeSb_6$ octahedra which form edge sharing chains along the $c$ axis. The space group is the centrosymmetric $Pnnm$ orthorhombic group \cite{petrovic03}, for which we report the factor group analysis in Table III. By using the correlation method \cite{fateley}, and after subtracting the acoustic modes, we find for the total irreducible representations:
\begin{equation}
\Gamma_{Pnnm}=A_g+2A_u+B_{1g}+B_{1u}+2B_{2g}+3B_{2u}+2B_{3g}+3B_{3u}.
\end{equation}

%<<<<<<<<<<<<<<<<<<<<<<<< TABLE >>>>>>>>>>>>>>>>>>>>>>>>>
\begin{table}[!ht]
\begin{center}
  \caption{\label{Pnnm} Normal modes analysis of $FeSb_2$ for the $Pnnm$ symmetry group.}
\begin{ruledtabular}
   \begin{tabular}{c|ccc|ccc}
Normal & Number & of  & modes & & &\\ 
modes & Total & Acoustic & Optical  & Raman& & IR\\
\hline 
$A_g$ & 1 &  & 1 &  aa, bb, cc & & \\
$B_{1g}$ & 1 &  & 1 & ab & & \\
$B_{2g}$ & 2 &  & 2 & ac & & \\
$B_{3g}$ & 2 &  & 2 & bc & & \\
$A_u$ & 2 &  & 2 &  & & \\
$B_{1u}$ & 2 & 1 & 1 &  &  & E//c \\
$B_{2u}$ & 4 & 1 & 3 &  & & E//b\\
$B_{3u}$ & 4 & 1 & 3 &  & & E//a\\

\end{tabular}
\end{ruledtabular}
\end{center}
\end{table}
%<<<<<<<<<<<<<<<<<<<<<<<< table>>>>>>>>>>>>>>>>>>>>>>>>>

By comparing our data with the theoretical calculation for the number of infrared active phonon modes along the $b$ axis direction ($E\parallel b$), one can notice that our optical measurement suggests the presence of four instead of three phonon modes. Nevertheless, the resonance frequencies of the modes at 257 and 271 cm$^{-1}$ are quite close to each other. Therefore, optical spectra may reveal a slightly lower symmetry for $FeSb_2$ with respect to the one predicted by the $Pnnm$ group \cite{petrovic03}.

Along the direction perpendicular to the $b$ axis ($E\perp b$) three phonon modes are present at low temperature, while only two of them are still observable above 50 K. If one assumes the $Pnnm$ group symmetry for $FeSb_2$, our spectra would imply that the axis perpendicular to $b$ in our experiment corresponds to the crystallographic direction $a$. On the other hand, the disappearance of the phonon mode at 216 cm$^{-1}$ at high temperatures might suggest a phase transition into a state, where this specific phonon is silent and not infrared active. Nevertheless, no structural changes have been found so far in $FeSb_2$ down to 10 K (Ref. \onlinecite{petrovic05}). 

\subsubsection{Fano analysis}

A blow up of the spectral range pertinent to the asymmetric mode at 260 cm$^{-1}$ for $E\perp b$ at 10 K is displayed in Fig. 3a. It is easily recognized that the Fano mode is more appropriate than a simple Lorentz oscillator in order to reproduce the overall shape of this mode. The temperature dependence of the fitting parameters for this peculiar mode is displayed in Fig. 3b-e. The oscillator strength is associated to the effective transverse charge and has a non-trivial temperature dependence (Fig. 3b) since it first decreases with increasing $T$, passes through a minimum between 50 and 70 K (i.e. in correspondence with the MI transition temperature) and then increases again.  

%<<<<<<<<<<<<<<<<<<<<<<<< FIGURE>>>>>>>>>>>>>>>>>>>>>>>>>
\begin{figure}[tbp]
   \begin{center}
   %\resizebox{9.0 cm}{!}{\includegraphics{Figura2.epsf}}
    \leavevmode
    \epsfxsize=10cm \epsfbox {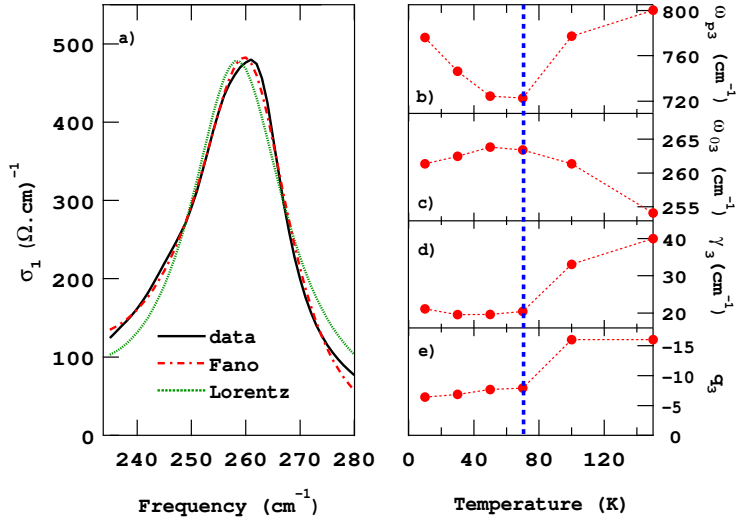}
     \caption{(color online) a) Blow-up of $\sigma_1(\omega)$ around the mode at 260 cm$^{-1}$ at 10 K and for $E\perp b$. The fits of this mode with either a Lorentz harmonic oscillator or a Fano lineshape are shown for comparison. b) Temperature dependence of the oscillator strength $\omega_{p3}$, c) resonance frequency $\omega_{03}$, d) scattering rate $\gamma_3$, and e) asymmetry factor $q_3$ of the mode at 260 cm$^{-1}$ within the Fano-like fit. The blue dashed line indicates the upper limit of T$_{MI}^{\perp b}$.}
\label{fig2}
\end{center}
\end{figure}
%<<<<<<<<<<<<<<<<<<<<<<<< figure >>>>>>>>>>>>>>>>>>>>>>>>>

A similar trend is also found for the resonance frequency $\omega_{03}$ (Fig. 3c), which slowly increases with increasing $T$ before displaying a well defined softening at high temperatures. X-ray diffraction studies \cite{petrovic05} have established that the $Fe-Sb-Fe$ bond angle, associated with the edge sharing octahedra along the $c$ axis, increases above 100 K. The $T$-dependence of the $Fe-Sb-Fe$ angle behaves rather closely to the behavior of $\omega_{03}$. These observations, together with the appearance at low $T$ of the mode at 216 cm$^{-1}$, are clear evidence for a structural distortion accompanying the MI transition. 
 
The damping factor $\gamma_3$ (Fig. 3d) increases above T$_{MI}^{\perp b}$. Since $\gamma\sim1/\tau$ , this indicates a two times higher phonon lifetime $\tau$ in the low $T$ insulating state. Such a strong correlation between the onset of the insulating state and the loss of phononic relaxational channels also suggests the presence of a coupling mechanism between the electronic degrees of freedom and the lattice vibrations, as already claimed for $FeSi$  (Ref. \onlinecite{damascelli97}).
 
Finally, Fig. 3e displays the $T$-dependence of the dimensionless Fano  parameter $q$. At high $T$, $|q|$ is of the order of 16. Such a high $q$-value implies that above 100 K an asymmetric phonon lineshape can be hardly recognized; a Lorentz harmonic oscillator would fit equally well this mode. On the other hand, $q$ takes values between -6.4 and -7.9 below T$_{MI}^{\perp b}$, indicating a sizable asymmetric lineshape. The presence of Fano lineshapes is predicted when a discrete lattice vibration is degenerate with states belonging to a continuum which may be of electronic or magnetic nature \cite{calvani01}. Interestingly, Fano lineshapes have been also observed in the infrared spectra of $FeSi$  (Ref. \onlinecite{damascelli97}). In that case, the asymmetry was ascribed to the interaction of the phonon mode with an electronic resonance peaked in the mid-infrared range, i.e. above the phonon resonance frequency. However, for $FeSb_2$ the sign of $q (<0)$ is consistent with the picture of the phonon at $\sim 260$ cm$^{-1}$ coupled to an energy scale which is lower than the phonon resonance frequency. The forthcoming analysis of the spectral weight distribution in the absorption spectrum of $FeSb_2$ will shed more light on this low energy scale. 

\subsection{Spectral weight analysis}

The $\sigma_1(\omega)$ spectra at 10  K present for both polarizations a deep minimum at about 280 cm$^{-1}$, which coincides with the onset of a broad and temperature independent tail, ascribed to the incoherent electronic contribution. We call such an energy, defining the onset of the interband transitions, $E_g$. $E_g\sim 280$ cm$^{-1}$ is therefore an appropriate estimation for the insulating optical gap in $FeSb_2$. The value of $E_g$ is sizably larger with respect to its estimation based on the activated behaviour of the resistivity, which indicates $\Delta_{\rho}(a,c)\sim 200$ cm$^{-1}$, and $\Delta_{\rho}(b)\sim 170$ cm$^{-1}$ (Ref. \onlinecite{petrovic03}). Similarly to the case of $FeSi$ (Ref. \onlinecite{paschen97}), this discrepancy in  the gap value among different experiments could be reconciled by assuming a small indirect energy gap for the transport properties and a larger direct gap for the optical response. On the other hand and at variance with $FeSi$ (Ref. \onlinecite{damascelli97}), $E_g$ does not display any temperature dependence up to the temperature ($T_{MI}$), at which the MI transition is observed. 

By integrating the optical conductivity (i.e., $\int_{0}^{\omega_c}\sigma_{1}^{Drude}(\omega)d\omega$), we can achieve the spectral weight encountered in the absorption spectrum over a frequency interval between zero and $\omega_c$. Because of the temperature dependence of $\sigma_1(\omega)$ (Fig. 2b), there is an obvious redistribution of the spectral weight from low to high frequencies in our spectra. A sum rule \cite{wooten,dressel} requires that the integrated area under $\sigma_1(\omega)$ must be unchanged at any temperature; therefore predicting the conservation of the spectral weight. In a conventional semiconducting scenario, the $T$-induced MI transition is simply due to the thermal excitation of charge carriers through the energy gap $E_g$. From an optical point of view, one would therefore expect that by lowering the temperature through $T_g$=$E_g$/$k_B$ the Drude term progressively vanishes while the spectral weight lost at $\omega<E_g$ redistributes just above and close to $E_g$. Such a picture is inadequate for $FeSb_2$. We find that the overall depletion of $\sigma_1(\omega)$ at 10 K leads to a removal of spectral weight at low frequencies, which mostly (i.e., 90 $\%$) piles up in the spectral range between $E_g$ and 3000 cm$^{-1}$. However, a full recovery of the spectral weight only occurs when integrating $\sigma_1(\omega)$ up to 1 eV. The observation, that one needs to go to unusually high frequency to satisfy the conduction sum rule, might suggest, similarly to $FeSi$ \cite{schlesinger93}, that the physics of $FeSb_2$ involves an energy scale much larger than the gap energy $E_g$. The transfer of spectral weight in spectroscopies of correlated electron systems has been discussed in the framework of the periodic Anderson model for the situation pertinent to the Kondo insulators \cite{rozenberg}.

%<<<<<<<<<<<<<<<<<<<<<<<< FIGURE>>>>>>>>>>>>>>>>>>>>>>>>>
\begin{figure}[h]
   \begin{center}
   %\resizebox{9.0 cm}{!}{\includegraphics{Figura3.epsf}}
    \leavevmode
    \epsfxsize=10cm \epsfbox {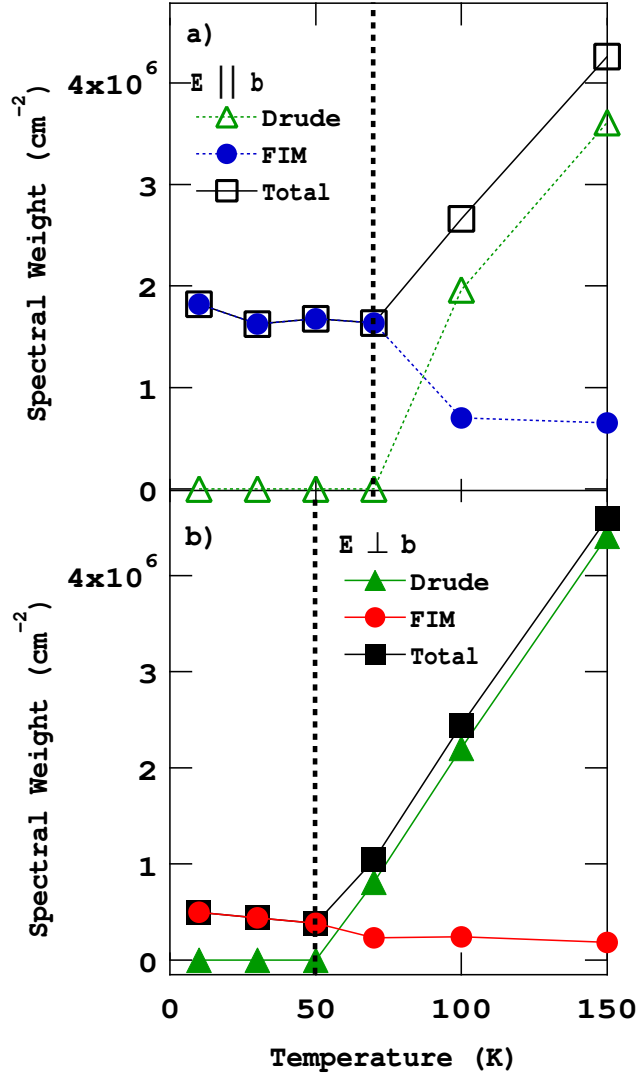}
     \caption{(color online) Temperature dependence of the spectral weight encountered in $\sigma_1(\omega)$ at $\omega<E_g$ for $E \parallel b$ (a) and $E \perp b$ (b). The black squares represent the integrals of $\sigma_1(\omega)$ up to 280 cm$^{-1}$ after having subtracted the fit to the phonon modes. The green triangles represent the spectral weight of the metallic Drude term. The blue and red circles represent the spectral weight ($\delta A$) of the broad infrared mode (FIM) for $E\parallel b$ and $E\perp b$, respectively.}
\label{fig3}
\end{center}
\end{figure}
%<<<<<<<<<<<<<<<<<<<<<<<< figure >>>>>>>>>>>>>>>>>>>>>>>>>

We can now calculate the total spectral weight encountered in $\sigma_1(\omega)$ of $FeSb_2$ at frequencies lower than $E_g$. To do so we have subtracted the phonons' contribution. The temperature dependence of this latter quantity is plotted in Fig. 4 for both polarization directions (open and filled black squares for $E\parallel b$ and $E\perp b$, respectively). We remark that at $T>T_{MI}$ a fraction of the total spectral weight below $E_g$ is due to a background of weight distributed over an interval of energies extending from 0 up to $E_g$. We associate such a background of spectral weight to the presence of a continuum of low lying excitations. This continuum of excitations defines a broad far infrared mode (FIM). The total spectral weight up to $E_g$ decreases with decreasing temperature down to $T_{MI}$ (Fig. 4). Below $T_{MI}$, some residual spectral weight (which can not be of vibrational origin) survives inside of the gap.
As demonstrated in Fig. 4, the spectral weight of the Drude term disappears below $T_{MI}$. The removed Drude weight is transferred at high energies and partially piles up in the spectral range covered by FIM. The size of this FIM term increases indeed with decreasing temperature, as shown in Fig. 4. The continuum of excitations, defining FIM, is better highlighted in the insets of Fig. 5. The red and blue area for for $E\parallel b$ and $E\perp b$, respectively, represent the spectral weight encountered in FIM (for the data at 10 K). Such areas ($\delta A$) are calculated by subtracting the spectral weight encountered in the Drude-Lorentz fit from the total spectral weight underneath the measured $\sigma_1(\omega)$ curve. Figure 5 displays then the temperature dependence of $\delta A$, normalized by its value at 10 K (Ref. \onlinecite{note}). At approximately $T_{MI}$ the spectral weight associated to FIM suddenly increases and the $T$-dependence $\delta A$ vaguely mimics the behaviour of an order parameter.

%<<<<<<<<<<<<<<<<<<<<<<<< FIGURE>>>>>>>>>>>>>>>>>>>>>>>>>
\begin{figure}[h]
   \begin{center}
   %\resizebox{9.0 cm}{!}{\includegraphics{Figura3.epsf}}
    \leavevmode
    \epsfxsize=10cm \epsfbox {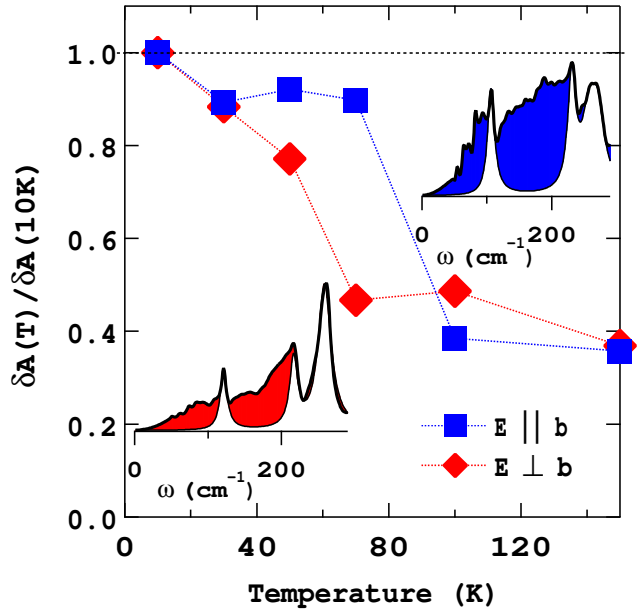}
     \caption{(color online) Temperature dependence of the normalized spectral weight ($\delta A$) of FIM. Insets show the area $\delta A$ (blue and red for $E\parallel b$ and $E\perp b$, respectively) at low temperature, encountering the spectral weight of FIM (see text).}
\label{fig4}
\end{center}
\end{figure}
%<<<<<<<<<<<<<<<<<<<<<<<< figure >>>>>>>>>>>>>>>>>>>>>>>>>

The temperature dependence of $\delta A$ seems to shape the intrinsic electronic properties of $FeSb_2$. In fact, the presence of such a (FIM) continuum implies that several localized states are present at low temperatures between the valence and the conduction band. Above $T_{MI}$ those states, generating FIM, are thermally occupied. Envisaging an overlapping of these states, a coherent electronic transport can then develop. In this context, it is not surprising that $T_{MI}$ is lower than $E_g$. Furthermore, the presence of this continuum of excitations supplies a natural explanation for the Fano behavior of the mode at $\omega_{03}$=260 cm$^{-1}$ for $E\perp b$. As anticipated above, the negative sign of the $q$ Fano parameter implies an interaction of the mode at $\omega_{03}$ with a continuum covering an energy spectral range located below $\omega_{03}$. Such a continuum of excitations is precisely provided by FIM. As comparison, we note that in the case of $FeSi$, where no evidence for an equivalent FIM feature was found, the asymmetry of the phonons was on the contrary ascribed to the red-shift of the gap edge \cite{damascelli97}. The gap edge in $FeSb_2$ is temperature independent. Therefore, this excludes any possibility for a temperature dependent coupling between the phonon modes and mid infrared excitations.

\section{Conclusion}
The electrodynamic response of $FeSb_2$ is characterized by a metal-insulator transition occurring at $T<70$ K, and opening of a gap $E_g\sim 280$ cm$^{-1}$. With decreasing temperature there is a depletion of $\sigma_1(\omega)$ signalling the removal of spectral weight, which is shifted to high energies up to 1 eV. Furthermore, we have observed that the suppressed Drude weight with decreasing temperature (i.e., at $T<T_{cr}$) partially piles up in the broad far infrared mode (FIM), which we have ascribed to a continuum of excitations. An ordinary semiconductor picture cannot account for the charge dynamic of $FeSb_2$. While some features in the optical response may be compatible with a Kondo insulator scenario, it remains to be seen how the presence of the continuum of excitations in the far infrared and at $\omega<E_g$ might eventually be consistent with such a description. Our findings challenge indeed the theoretical treatment of the Kondo insulators within the framework of the periodic Anderson model \cite{rozenberg}.

\acknowledgments
The authors wish to thank J. M\"uller for technical help, and G. Caimi, Z. Fisk and P.C. Canfield for fruitful discussions. This work has been supported by the Swiss National Foundation for the Scientific Research, within the NCCR research pool MaNEP and by the Office of Basic Energy Sciences of the U.S. Department of Energy and it was partly carried out at the Brookhaven National Laboratory, which is operated for the U.S. Department of Energy by Brookhaven Science Associates (DE-Ac02-98CH10886).

\newpage


\begin{thebibliography}{10}
\bibitem{note}
The purpose of the normalized representation of $\delta A$ is to better emphasize the $T$-dependence of the FIM spectral weight, already shown in Fig. 4, and to stress the possible order parameter behaviour of this quantity.
\end{thebibliography}
\end{document}